\documentclass[aps,prl,showpacs,superscriptaddress]{revtex4}
\usepackage{amsmath}
\usepackage{epsfig}
\usepackage{amssymb}
\usepackage{dcolumn}
\usepackage{graphicx}
\usepackage{dcolumn}

\usepackage{color}

\newcommand{\beq}{\begin{equation}}
\newcommand{\eeq}{\end{equation}}

\begin{document}
\date{\today}

\title{Bright solitons in non-equilibrium coherent quantum matter}

\author{F. Pinsker}
\affiliation{Clarendon Laboratory, University of Oxford, Parks Road, Oxford OX1 3PU, United Kingdom}
\email{florian.pinsker@physics.ox.ac.uk}

\author{H. Flayac}
\affiliation{Institute of Theoretical Physics, \'{E}cole Polytechnique F\'{e}d\'{e}rale de Lausanne EPFL, CH-1015 Lausanne, Switzerland.}
\email{hugo.flayac@epfl.ch}

\begin{abstract}

We theoretically demonstrate a mechanism for bright soliton generation in spinor non-equilibrium Bose-Einstein condensates made of atoms or quasiparticles such as polaritons in semiconductor microcavities.  We give analytical expressions for bright (half) solitons as minimizing functions of a generalized non-conservative Lagrangian elucidating the unique features of inter and intra-competition in non-equilibrium systems. The analytical results are supported by a detailed numerical analysis that further shows the rich soliton dynamics inferred by their instability and mutual cross-interactions.

\end{abstract}

\maketitle

\section{Introduction} Investigations on Bose-Einstein condensation (BEC) have been key in understanding quantum matter at very low temperature highlighting its distinctive coherent wave properties \cite{leg,Stri}. BEC is a macroscopic quantum phenomenon often involving a large number of interacting coherent quantum particles \cite{Stri}. In the last two decades the phase transition of many-body systems to a BE condensed state has become feasible within a great variety of different physical systems such as solid-state light-matter systems, various species of atoms and molecules and for gases of photons or classical waves \cite{photon,Light, exit, Fleisch}. Among the most intriguing  properties of all these condensed coherent matter states are regimes of frictionless flow (superfluidity) \cite{Lond} below a critical velocity \cite{heat, Pin4} or the response to motion via elementary excitations such as quantized vortex rings, quantum vortices and lattices thereof, dark and bright solitons \cite{Lago, Pin1,Pin2, Pin3,Pin4,Pin6,Pin7,Tris}. Different kinds of BEC inherit properties depending on the nature of the entities constituting them, such as variable inter-particle interactions due to Feshbach resonances when applying certain external fields in atomic and polariton condensates \cite{fesh, fesh2} or local particle sources and sinks in non-equilibrium condensates of exciton-polaritons \cite{Light, Deng} or atoms \cite{AtomLaser}. In turn the out of equilibrium aspect of open quantum systems as well as the nature of interactions between the coherent particles is key for the possible pattern formations in those condensates and thus the properties of excitations within the condensate \cite{Light, Stri, Pin6}.

Attractive Bose-Einstein condensates of atoms were experimentally realized \cite{fuck} as well as repulsive condensates of polaritons \cite{exit}. Later it was shown that self-interactions of condensed atoms or polaritons can be tuned continuously via utilizing Feshbach resonances \cite{fesh, fesh2} providing one possible route towards implementing attractive non-equilibrium condensates. In addition quasiparticles in solid state systems can have an effectively negative mass proportional to the inverse of the second derivative of their dispersion, for which polaritons are an example, and so the mathematical form of the mean field theory corresponds to effectively attractive self-interactions providing an alternative approach to Feshbach resonances \cite{Sich,mk}. Polaritons are quasiparticles within semiconductor microcavities which satisfy Bose-Einstein statistics \cite{Light} and show macroscopic occupation of the lowest energy state with off-diagonal long range order \cite{exit}. Temporal coherence is limited to a few $ps$ due to the continuous loss of particles and information \cite{loss}. Excitons are coupled electron-hole pairs of oppositely charged spin-half particles in a semiconductor bound by the Coulomb interactions \cite{Deng}. Since states of excitons interact with light \cite{Hopf} in the strong coupling regime exciton-polaritons (polaritons) are formed \cite{Weis} and as polaritons are $10^9$ times lighter than rubidium atoms \cite{exit}, condensation can be observed on the Kelvin range in CdTe/CdMgTe/GaAs micro-cavities \cite{exit,Light, Ohadi} and even at room temperatures in GaN or flexible polymer-filled micro-cavities \cite{Christopoulos,Plum}. Polaritons are a prime example for an open quantum system and allow to test non-equilibrium properties and are analoguous in their mathematical structure to open atomic BEC systems when considering mean field properties of their coherent ground states \cite{AtomLaser}.

The system we shall study in this paper is an effectively attractive spinor non-equilibrium condensate within a mean-field parameter regime and with potential realization in non-equilibrium systems of atoms, molecules or quasiparticles such as polaritons. A key requirement for the applicability of our theory is the continuous exchange of particles with an external bath, which allows richer dynamical behavior of the spinor field and the emergent excitations within. Effective mechanisms to generate excitations in general BEC within the Gross-Pitaevskii (GP) mean-field parameter regime \cite{Light, Stri} are important tools for investigations on their properties both experimentally and theoretically \cite{firstsol} and carry the possibility for applications e.g. in optical computing \cite{me,Tris}. Vortices with quantized circulation in superfluids are a key example of a topologically stable excitation in a $2$d BEC system guided by a nonlinear Schr\"odinger-type equation. Recently, they have been predicted \cite{Rubo,FlayacHV,KeelingV} and observed \cite{LagoudakisHV} in polariton condensates. Their counterparts in $1$d repulsive condensates are solitons, localized excitations moving without deformation over time \cite{Pin6}. They have been predicted in several contexts \cite{Pin6, Monopole1, Monopole2, KivsharDark, decay, Dreis, CuitiSol,FlayacHS} and observed so far mostly in planar cavities \cite{AmoS,HivetHS,Sich}. In addition state-of-the-art investigations describe driven dissipative bright solitons under non-resonant excitation \cite{OstrovskayaDS,SavenkoSesam} for the repulsive BE condensate. In this work we provide an analytical mean field non-equilibrium theory for bright solitons and predict numerically the {\it bright soliton} patterns that emerge within an effectively attractive spinor condensate. Our numerical results are supported by analytical parameter equations and particularly explicit expressions for the stationary solutions are obtained by this method.

The paper is organized as follows. First we introduce the state-of-the-art mean field model. Then we simplify the mean-field model to the Keeling/Berloff form of the governing non-conservative partial differential equation (PDE). Based on this we introduce the Lagrangian treatment of the PDE for a bright soliton solution. Subsequently we extend the considerations to the spin sensitive case. Finally we turn to numerical investigation of the bright solitons and particularly compare numerics with analytics.

\section{The model} We now turn to the state equation for a two component spinor field $\boldsymbol{\psi}=(\psi_+,\psi_-)^T$ that describes the two coherent macroscopically occupied many body states $\psi_\pm$ where we refer to $\pm$ as spin although $\pm$ can be thought of as any two coherent hyperfine states in the considered many body system. The system/scenario of interest is modeled by $1$D Ginzburg-Landau type equations coupled to a rate equation for the reservoir $n_R$ capturing the non-condensed particles within the bath feasible in open systems such as polaritons \cite{Pin6} or atoms \cite{Stri, AtomLaser},
\beq\label{system}
 i \partial_t  \psi_\sigma =  \frac{\mathcal H_{\sigma}}{\hbar} \psi_\sigma
\eeq
with $\sigma \in \{+,-\}$ and
\begin{equation}\label{natural}
 \frac{\mathcal H_\pm}{\hbar}  \psi_\pm = (1- i \eta) \bigg( - \frac{\hbar \nabla^2}{2 m} + \alpha_1 ( |\psi_\pm|^2 + n_R) + \alpha_2 |\psi_\mp|^2  +  U  \bigg) \psi_\pm +    \frac{i }{2} \left(\gamma  n_R - \Gamma  \right) \psi_\pm
 \end{equation}
Here $\eta$ is the amount of energy relaxation of the condensed phase, $\alpha_1$ denotes the same spin interaction strength and $\alpha_2$ the cross spin interactions. $m$ is the effective mass of the particles within the condensate phase. Furthermore we denote the pumping rate into the condensate $\gamma$ and the decay rate for condensed particles as $\Gamma$. Pumping is modeled by a spatial and time dependent function $P$ which induces particles into the reservoir/bath of the system \cite{Pin6, AtomLaser},
\beq\label{Res}
\partial_t n_R   = P -  n_R (\Gamma_R + \gamma ( |\psi_+|^2+ |\psi_-|^2))
\eeq
where \cite{AtomLaser} has been extrapolated to the spin sensitive case. The reservoir balance equation is fixed at a decay rate $\Gamma_R$ and driven by the spatially dependent pump term $P=A_P \exp(-x^2/\sigma_P^2)$ where $\sigma_P$ is the width, and $A_P$ is the amplitude modeling the scattering into the ground state. The reservoir scatters particles into the condensate at a rate $\gamma$. The coupled system of equations describes the time evolution of the condensate and its reservoir and is an example for a non-resonant pumping scheme in polariton condensates \cite{Wout,Pin6} with analog in atoms \cite{AtomLaser}. For atoms one can understand the spinor field as the two hyperfine states, say A and B, of a BEC $\boldsymbol{\psi}$ \cite{Stri}. Now, let us turn to analytical considerations for the spin coherent case.

\section{Attractive non-equilibrium condensate} Let us first consider a simple case to introduce the analytical arguments presented in the following. In the case of spin coherence we can simplify the system of equations \eqref{system} to \cite{Wout, KeelingV}
\begin{equation}\label{natural2}
 \frac{i}{\hbar} \partial_t  \psi = (1- i \eta) \bigg( - \frac{\hbar \nabla^2}{2 m}  + \alpha'_1 ( |\psi|^2 + n_R)   + U \bigg) \psi +  \frac{i }{2} \left(\gamma  n_R - \Gamma  \right) \psi.
\end{equation}
In addition we note that the true form of the non-equilibrium dynamics has not yet been measured experimentally and both models \cite{Wout, KeelingV} are often used to successfully describe experiments with polariton condensates. The actual form of the density saturation in \eqref{natural2} however,  becomes relevant for large amplitudes of the wave function, i.e. large occupation numbers of the condensed phase. Both phenomenological mean field models are connected by Taylor expanding the pumping term stated in \cite{Wout} to the first order as outlined in \cite{Tris}.
Now, starting from \eqref{natural} we approximate/simplify the complex terms of the condensate wave equation \eqref{natural2} by the reservoir and decay of particles as in \cite{ Dreis, Tris},
\beq\label{ooo}
  \frac{i }{2} \left(\gamma  n_R - \Gamma  \right) \simeq i ( \delta + \varepsilon \alpha'_1 |\psi|^2),
\eeq
to obtain a mathematical form easily accessible to apply the following analytics.
For attractive condensates the coefficient of the real-valued nonlinear term $\alpha'_1$ in \eqref{natural2} is negative, while it is positive when modeling repulsively interacting condensates. Bright soliton are solutions for the GPE equation with negative nonlinearity and thus excitations in attractive BE condensates. In \cite{1,Non,Polo} the bright soliton solutions for the complex cubic-quintic Ginzburg-Landau equation have been considered. By taking \eqref{ooo} into account eq. \eqref{natural2} can be regarded  as a special case of this cubic-quintic  PDE, but without quintic terms and we can apply and later extend the analysis outlined in \cite{1,Non,Polo}. However, in contrast to their derivation we assume that the sign of the linear complex term is positive, while stability requires us to introduce a cutoff \cite{KeelingV} and to set in turn $\varepsilon = - |\varepsilon|$ which is the magnitude of the density dependent decay/saturation of condensed particles. Further we neglect the blueshift from the reservoir (which can be absorbed into the phase) and consider no external potential, i.e. $U=0$ in \eqref{natural2}. Rescaling $\vec r = \sqrt{2m/\hbar} \cdot \vec x$, $z = \hbar t$ and $\psi \to \frac{1}{\sqrt{\alpha'_1}} E$ yields the governing PDE for the non-equilibrium system,
\begin{equation}\label{natural3}
i \partial_z  E = - \Delta E  \pm  |E|^2 E + i \delta E + i \varepsilon |E|^2 E
\end{equation}
with the d dimensional Laplacian $\Delta E = r^{1-d} \partial_r (r^{d-1} \partial_r E)$. In the following we assume $d=1$ as we are interested in the description of bright solitons, but write down the Lagrangian procedure for general dimensions. For negative self-interactions \eqref{natural3} with $1/2$ in front of the Laplacian and with $\delta =0$ and $\varepsilon =0$ the equation possesses an exact bright soliton solution \cite{Pin4,ching}
\beq\label{anal}
E(z,t) = A \cdot {\rm sech}(A( z - v t)) \exp(i (\kappa z - \omega t)),
\eeq
where $A$ denotes the amplitude of the excitation, $\kappa$ the soliton wave number $\omega = (\kappa^2- A^2)/2$ is the soliton dispersion relation and $v = \partial \omega / \partial \kappa = \kappa$ the soliton velocity \cite{ching}. Later we extend \eqref{anal} to the variational ansatz we will use to describe non-equilibrium bright solitons. The Lagrangian of our system \eqref{natural2} denoted $\mathbb L$, can be separated additively into its conservative part $\mathbb L_C$  (real-valued energy terms)  and its dissipative part, inducing the non-equilibrium properties of the condensate wave function due to pumping and loss of the particles, the perturbation $\mathbb L_Q$. Thus the total Lagrangian is given by
\beq\label{laga}
\mathbb L = \mathbb L_C + \mathbb L_Q
\eeq
 with components
\begin{equation}
\mathbb L_C =  r^{d-1} \left( \frac{i}{2} \left(E \partial_z E^* - E^* \partial_z E \right) + |\nabla E|^2 - \frac{|E|^4}{2} \right)
\end{equation}
and
\beq
\mathbb L_Q = i r^{d-1} \left( \delta |E|^2+ \varepsilon \frac{|E|^4}{2}  \right).
\eeq
We want to point to the analysis presented in \cite{Non, method,Polo} for comparable applications of the non-conservative Lagrangian method with generalized forces in the context of Ginzburg-Landau type or complex Gross-Pitaevskii-type equations. We take \eqref{laga} and apply Hamilton's principle of vanishing variation, i.e. $\delta (\int \int (\mathbb L_C + \mathbb L_Q) dz dr) = 0 $, which is obtained for the minimizing condensate wave function (minimizer). The exact minimizer is a solution of the Euler-Lagrange equations, obtained from the Lagrangian $\mathbb L$ of the system by applying
\beq
\sum_\eta \frac{d}{d \eta} \left( \frac{\partial}{\partial E^*_\eta} \mathbb L - \frac{\partial }{\partial E^*} \mathbb L \right) = 0.
\eeq
Although we do not know the exact form of the minimizer we can approximate the minimizer by an ansatz similar to the bright soliton solution for equilibrium condensates \eqref{anal}, the results will the later be compared to numerics yielding the numerically generated minimizer justifying our ansatz. To proceed this way we first optimize over the transverse coordinates \cite{Non2,Non}
\begin{equation}
\int dr \left( \sum_\eta \frac{d}{d \eta} \left( \frac{\partial}{\partial E'^*_\eta} \mathbb L - \frac{\partial }{\partial E^*} \mathbb L \right)  \right) = 2 {\rm Re} \int dr \left(  \frac{\partial }{\partial E^*} \mathbb L - \sum_\eta \frac{d}{d \eta} \frac{\partial}{\partial E'^*_\eta} \mathbb L \right)
\end{equation}
with $\eta =r,z$ where the r.h.s. corresponds to the non-conservative part \cite{Non2}. Furthermore by denoting $L_c = \int dr \mathbb L_c$ we effectively consider \cite{Non2}
\begin{equation}\label{into}
\frac{d}{dz} \left(\frac{\partial L_c}{\partial \nu'}  \right) - \frac{\partial L_c}{\partial \nu} = 2 {\rm Re} \int r^{d-1} dr \left(i \delta E + i \varepsilon |E|^2 E \right) \frac{\partial E^*}{\partial \nu},
\end{equation}
where the generalized forces on the r.h.s. represent the non-equilibrium processes perturbing the condensate.
Here $\nu = A(z), W(z), P (z), C(z)$ for the nonlinear bright soliton ansatz of the condensate wave function representing an approximate minimizer of \eqref{laga} that is given by
\begin{equation}\label{Ansatz}
E(z,r) =A(z) \cdot {\rm sech} \left(\frac{r}{W(z)} \right) \exp(i P (z) + i C(z) r).
\end{equation}
We insert this ansatz in $L_c$ and subsequently plug the expression for $L_c$  into the l.h.s of eq. \ref{into}. Then variation over $A$ yields
\begin{equation}
P' = \frac{-1 + 2 A^2 W^2 - 3 C^2 W^2}{3 W^2}
\end{equation}
utilizing here the simple notation $' = \partial_z$. By variation regarding $W$ we obtain
\beq
W = \pm \frac{1}{\sqrt{-A^2 + 3 C^2 + 3 P'}}
 \eeq
and variation regarding $P$ yields
\beq
\partial_z ( A^2 W) = 2A^2 \delta W + \frac{4}{3} A^4 \varepsilon W.
\eeq
Finally variation regarding $C$ gives us the formula
\beq\label{C}
4 A^2 C W = 0.
\eeq
Expression \eqref{C} implies that $C = 0$ as long as we have a non vanishing amplitude and do not consider the homogeneous solution.
We note that our results are consistent with \cite{method, Polo, Non} and point out that the sister case of dark solitons in non-equilibrium BEC has been considered in \cite{Non}. For vanishing time dependence of the density profile, i.e. $A$ and $W$ are stationary, we obtain a simple formula for the amplitude of the bright soliton in terms of the linear pumping (reduced by the linear decay) $\delta$ of polaritons/atoms into the condensate phase and the density dependent decay of strength $\varepsilon = - |\varepsilon|$,
\beq\label{A}
A = \pm  \sqrt{\frac{3}{2} \frac{\delta}{|\varepsilon|} }.
\eeq
One observes in $\eqref{A}$ that the amplitude is higher for increased linear absolute gain $\delta$ and lower for increased density dependent decay $\varepsilon$. The simple limit $\delta \to 0$, yields a vanishing occupation of the condensate phase, because $A \to 0$ consistent with experiments.
For the width of the bright soliton we obtain
\beq
W = \pm \frac{\sqrt{2}}{A} = \pm \sqrt{\frac{4}{3} \frac{|\varepsilon|}{\delta} },
\eeq
which differs by a factor of $\sqrt{2}$ from the corresponding dark soliton expression \cite{Non}. As the amplitude of the condensate increases, the width of the bright soliton decreases, due to this reciprocial relation. Finally the phase of the non-equilibrium bright soliton satisfies,
\beq
P(z) = \frac{A^2}{2} z + const = \frac{3}{4} \frac{\delta}{|\varepsilon|} z + const,
\eeq
and increases quadratically with amplitude.

\section{Attractive non-equilibrium spinor condensate} While we have introduced the non-conservative Lagrangian method for solitons in attractive BEC, next we consider the complete spinor system \eqref{system} for one spatial dimension \cite{Pin6} and formally proceed similarly as in \cite{Non}. Using the Keeling/Berloff non-equilibrium terms for the governing condensate \eqref{ooo} we generalize them for the spinor condensate to
\beq\label{ooo2}
  \frac{i }{2} \left(\gamma  n_R - \Gamma  \right) \simeq i ( \delta + \varepsilon \alpha'_1 (|\psi_+|^2+|\psi_-|^2)),
\eeq
which is the mathematical form we shall analyst in the following. By applying analogous rescaling as in the coherent case  \eqref{natural3} we obtain the spinor PDEs
\begin{equation}\label{natural4}
i \partial_z  E_\pm = - \Delta E_\pm  -  (|E_\pm|^2 + \gamma  |E_\mp|^2) E_\pm + i \delta E_\pm + i \varepsilon (|E_\pm|^2 + \gamma_2 |E_\mp|^2) E_\pm,
\end{equation}
fully describing state-of-the art polariton experiments in appropriate regimes and which are analogous to atomic BEC mean field models \cite{Stri}.
Here $\gamma = \alpha_2/\alpha_1$ is the magnitude of cross-scattering between the two spin components and $\gamma_2$ models misalignment between self- and cross-spin decay. Correspondingly, the non-conservative Lagrangians (for dissipative and conservative part) for the attractive spinor condensate are
\begin{equation}
\mathbb L_C = \sum_{k \in \{+, - \}}  \bigg( \frac{i}{2} \left(E_k \partial_z E^*_k - E^*_\pm \partial_z E_k \right) + |\nabla E_k|^2  - \frac{|E_k|^4}{2}  \bigg) -  \gamma  |E_{+}|^2 |E_-|^2,
\end{equation}
 including  cross-interaction terms in contrast to the coherent states formalism above, and
\begin{equation}\label{last}
\mathbb L_Q =\sum_{k \in \{+, - \}} i \bigg( \delta |E_k|^2+  \varepsilon \frac{|E_k|^4}{2}  \bigg) + i \varepsilon \gamma_2  |E_{+}|^2 |E_-|^2.
\end{equation}
Here the last term in \eqref{last} is due to interspecific competition between the two spin components.
We adapt ansatz \eqref{Ansatz} to the spinor case,  writing $\Psi' = (E_+,E_-)$ we define its two components by
\begin{equation}\label{a22}
E_\pm(z,t) =   A_\pm(z) \cdot \exp \left(- \frac{(r \pm r_0)^2}{W_\pm(z)} \right) \exp(i P_\pm (z) + i C_\pm (z) r),
\end{equation}
where we model the bright solitons in terms of a gaussian instead of the ${\rm sech} (x)$ discussed in the previous section. The half-bright solitons are parametrized through the time dependent  amplitude $A_\pm$, width $W_\pm$, phase $P_\pm$, spatially dependent phase factor $C_\pm$ and a distance between the half-solitons given by $r_0$.
This ansatz describes two half-bright solitons positioned at $r = \pm r_0$ correspondingly. To integrate the cross-interaction term we proceed as follows. First we obtain the integrated conservative part of the  Lagrangian as in the previous section. The result is
\begin{equation}\label{conscoup}
L_c = \sum_{\sigma = +,-}\frac{\sqrt{\pi}}{4} A^2_\sigma \sqrt{\frac{1}{W_\sigma}} \left(-A^2_\sigma W_\sigma +
   2 \sqrt{2} (1 + W_\sigma (C^2_\sigma + P'_\sigma)) \right)  - \gamma A_+^2 A_-^2 \sqrt{\frac{\frac{\pi}{2}}{\frac{1}{W_\sigma} + \frac{1}{W_{\overline \sigma}}}} \exp{ \left(- \frac{8 r_0^2}{W_\sigma +W_{\overline \sigma}} \right)}.
\end{equation}
When the distance between the bright solitons is $r_0 \to \infty$ the cross-interaction is effectively zero, which is plausible due to finite sized widths of the bright solitons. Similarly as $\gamma \to 0$ \eqref{conscoup} uncouples as well the conservative part of the two components and we consider the case of two independent half-solitons, which resemble the single component case.  In accordance with the treatment of the previous section the generalized integrated Lagrangian equations are
\begin{equation}\label{into2}
\frac{d}{dz} \left(\frac{\partial L_c}{\partial \nu'}  \right) - \frac{\partial L_c}{\partial \nu} =  2 {\rm Re} \int r^{D-1} dr \bigg(i \delta E_\sigma +  i \varepsilon (|E_\sigma|^2 + \gamma_2 |E_{\overline \sigma}|^2) E \bigg) \frac{\partial E^*}{\partial \nu}.
\end{equation}
We remark that to obtain complete decoupling between both spin components we consider the limit $\gamma \to 0$ and $\gamma_2 \to 0$. Here $\gamma_2 \to 0$ corresponds to the case of vanishing interspecific competition \cite{Non}. In general (for $\gamma > 0$ and $\gamma_2 > 0$)  component $+$ will impact the parameters of the bright soliton within the other component $-$ and vice versa, i.e. the system is coupled through cross-interaction and cross-competition.

Using \eqref{conscoup}, \eqref{into2} and the $8$ parameter ansatz \eqref{a22} we obtain a set of $8$ coupled ordinary differential equations for those parameters: Variation over $W_\pm$ yields
\begin{widetext}
\beq
P'_\pm =\frac{A^2_\pm}{2 \sqrt{2}} - C^2_\pm + \frac{1}{W_\pm} + \gamma \left(\frac{B^2 \sqrt{\frac{1}{W_\pm} + \frac{1}{W_\mp}} W^2_\mp}{
 \sqrt{\frac{1}{W_\pm}} (W_\pm + W_\mp)^2}  + \frac{16 A_\mp^2 \exp{ \left(- \frac{8 r_0^2}{W_\pm + W_\mp} \right)} r_0^2 \sqrt{\frac{1}{W_\pm} +\frac{1}{W_\mp} } W_\mp W_\pm^{3/2}}{(W_\pm + W_\mp)^3} \right),
\eeq
with $\gamma$ denoting the cross-interaction strength, and variation of $A_\pm$ gives
\beq
C_\pm =\pm \sqrt{ \frac{ \sqrt{2} A^2_\pm - 2 P'_\pm - \frac{2}{W_\pm}  + \gamma \exp{ \left(- \frac{8 r_0^2}{W_\pm + W_\mp} \right)} \frac{2 A^2_\mp \sqrt{\frac{1}{W_\pm}}}{\sqrt{\frac{1}{W_\pm} + \frac{1}{W_{\mp}}}}}{2}}.
\eeq
Finally variation of $P_\pm$ and $C_\pm$ yields correspondingly
\begin{equation}\label{A2}
 \partial_z \left(A^2_\pm \sqrt{W} \right) =  A_\pm^2 \left(\delta  \sqrt{W_\pm} + \frac{\varepsilon}{\sqrt{2}} \left( A_\pm^2 \sqrt{W_\pm} +  \frac{\gamma_2 \sqrt{2} A_\mp^2}{\sqrt{\frac{1}{W_\pm} + \frac{1}{W_\mp}}} \exp{ \left(- \frac{8 r_0^2}{W_\pm + W_\mp} \right)} \right) \right),
\end{equation}
 where $\gamma_2$ denotes the interspecific competition strength, and
\beq\label{C2}
A^2_\pm C_\pm \sqrt{2 \pi} \sqrt{W_\pm} =  \varepsilon \gamma_2  A_\pm^2  A_\mp^2 \left(- \frac{r_0 \sqrt{2 \pi} W_\pm}{\sqrt{\frac{1}{W_\pm} + \frac{1}{W_\mp}} (W_\pm + W_\mp)} \exp{ \left(- \frac{8 r_0^2}{W_\pm + W_\mp} \right)} \right).
\eeq

\end{widetext}
Let us now derive a stationary density solution, where $\partial_z A_\pm =0 = \partial_z W_\pm$. Assuming $r_0 =0$ the r.h.s of \eqref{C2} vanishes, so $C_\pm =0$ for a nontrivial non-homogeneous solution, which reduces the system to $6$ free parameters. Furthermore we suppose $W_\pm = W_\mp$. Using $\eqref{A2}$ we obtain for the amplitudes
\beq\label{sa}
A_\pm^2 = \frac{\sqrt{2}\delta}{\varepsilon} \left(\frac{-1}{1+\gamma_2} \right).
\eeq
Note that $\varepsilon = - |\varepsilon|$. Formula \eqref{sa} implies that the higher the interspecific cross-competition, the lower the amplitude of each component , while the dependence on $\delta$ and $\epsilon$ corresponds with the findings for the spin coherent case. The widths of the two bright solitons satisfy
\beq
W_\pm =  \left|- \frac{4 \varepsilon }{\delta}  \left(\frac{1+\gamma_2}{ 1+\gamma } \right) \right|,
\eeq
which again is in accordance with the findings made in the previous section. However here for larger inter and intra component decay the solitons widen and as pumping increases the bright solitons narrow. Finally the phases are given by
\beq
P_\pm (z) =  \frac{A^2 (1+ \gamma)}{4 \sqrt{2}} z + c_0  = - \frac{\delta}{4 \varepsilon }  \left(\frac{ \gamma +1}{1+\gamma_2} \right)  z + c_0,
\eeq
where $c_0$ is the constant of integration. Note that inserting these expressions into \eqref{a22} yields the two condensate wave functions of the spinor non-equilibrium system.

In the following section we show numerical phenomenology of bright solitons and spinor bright solitons within non-equilibrium condensates and compare some numerics with the above analytical results.

%
%

\section{Numerical Results}
\begin{figure}[!t]
\[
\]
\includegraphics[width=0.75\linewidth]{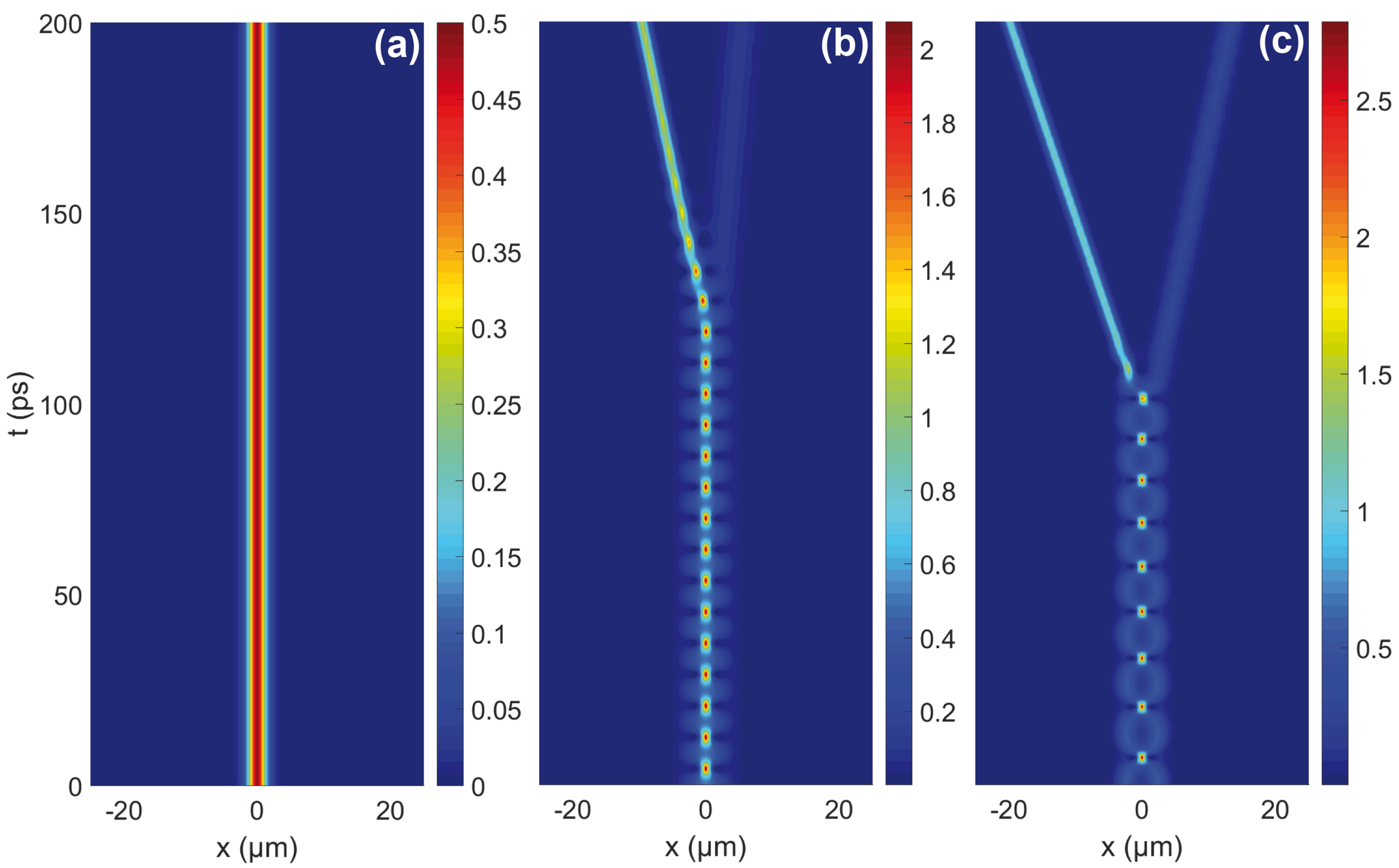}
\caption{Bright solitons in an equilibrium scalar condensate. The colormap shows the chemical potential $\mu(x,t) = \alpha_1 n(x,t)$ (meV). (a) Stable soliton solution from $\psi_b$. (b) Unstable breathing obtained imposing a too large intial condition. (c) Bright soliton attraction and oscillations.}
\label{Num1}
\end{figure}

In this section we present direct numerical solution to Eqs.(\ref{natural},\ref{Res}) that were computed with an adaptative step Runge-Kutta-based solver. We worked here with a spatial resolution of $h=0.25 \mu$m.
Let us first recall the case of a conservative closed condensate (e.g. an atomic condensate without exchange of particles or a polariton condensate with very long lifetimes due to high quality cavities). This setting corresponds to the assumptions $P=0$ and $\Gamma=0$ in Eqs.(\ref{system},\ref{Res}). In the coherent component case, which requires to set in addition $\alpha_2=0$, the systems reduces to considering the single component evolution, since $\psi_+ = \psi_- =\psi$. It is well-known that bright solitons or rogue waves naturally emerge in attractive condensates from the collapse of an inhomogeneous particle distribution \cite{Pin4}. We follow this study and let us consider in the remainder of this work typical exciton-polariton condensates parameters to be as explicit as possible and without loss of the generality for the observations made here.  In these systems the temporal, spatial and energy scales are respectively a few $ps$, $\mu m$ and $meV$s. Typical scenarios of different development in time of the bright soliton is shown in Fig. \ref{Num1}: panel (a) shows the unchanged form of the wave function evolved as described above starting from an initial bright soliton wavefunction $\psi_b(x)=\sqrt{n_0}{\rm{sech}}(x/\xi)$ where $\xi=\hbar/\sqrt{m \mu}$ is the healing length setting the soliton size and $\mu=\alpha_1 n_0$ is the chemical potential.  Panel (b) shows a realization where the initial wavefunction is set twice the size, i.e. with a spatial extension of 2$\xi\simeq2\mu$m. Here the interactions are not properly compensated by dispersion and the wavepacket/bright soliton starts collapsing. Subsequently it expands again when dispersion becomes dominant forming a clear {\it breathing pattern} that demonstrates their intrinsic instability due to the attractive interactions. Around $120$ ps after starting the process the finite numerical tolerance is enough to spontaneously break the symmetry. Then the bright soliton escapes to w.l.o.g. the left transforming the excess of energy into motion. Finally in the panel (c) we have initially prepared two bright solitons separated by $4$ $\xi$ and we observe their expected attraction. Indeed density maxima are bound to attract each other when $\alpha_1<0$. The attraction infers a velocity to the solitons that cross each other and repel again performing dipolar oscillations until they split at $t\simeq100$ ps.

\begin{widetext}
\begin{figure*}[!t]
\includegraphics[width=1\linewidth, height=0.4\linewidth]{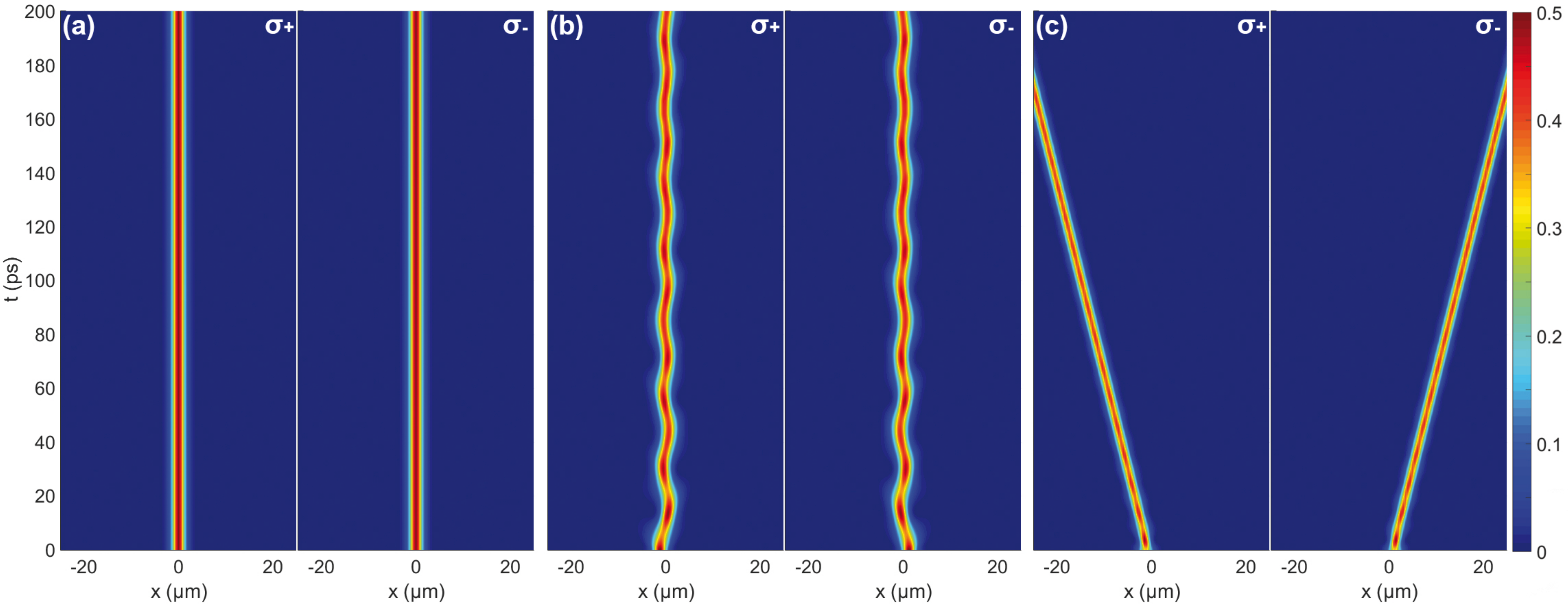}
\caption{Bright solitons in an equilibrium spinor condensate, the colormap shows the chemical potential. (a) Stable soliton solution from $\psi_\pm=\psi_b$. Half-solitons (a) attraction for $\alpha_2<0$ and (c) repulsion for $\alpha_2>0$.}
\label{Num2}
\end{figure*}
\end{widetext}

Now we consider the equilibrium \emph{spinor} case where both components are taken into account. In correspondence with above findings, in the case of negligible cross interactions, where $\alpha_2=0$ a stable spatial soliton in both component can be set via $\psi_\pm=\psi_b$ as shown in Fig. \ref{Num2} (a). Here the left (right) part shows the $\sigma_+$ ($\sigma_-$) component dynamics. Now if the solitons in each component (half-solitons) are spatially separated, they attract for $\alpha_2<0$ and repel when $\alpha_2>0$ as shown in Fig.\ref{Num2} (b) and (c) respectively as opposed to the dark soliton case \cite{Monopole2}. First note that in contrast to the dark soliton case where the effective magnetic field was shown to accelerate the half-solitons appearing on top of a nonzero density background, here it simply results in a pseudospin precession namely a weak population exchange between the two components. Secondly note that the systems chemical potential is modified as $\mu=n_0(\alpha_1 + \alpha_2)/2$ in the spinor case.

These numerics provide us intuition for the non-equilibrium case and we can already imagine that the competition between the intra-component repulsion (Fig. \ref{Num1}) and cross interactions (Fig. \ref{Num2}) combined with the breathing effect can lead to rich dynamics of the non-equilibrium spin sensitive BE condensate.

\begin{figure}[!t]
\includegraphics[width=0.95\linewidth, height=1.1\linewidth]{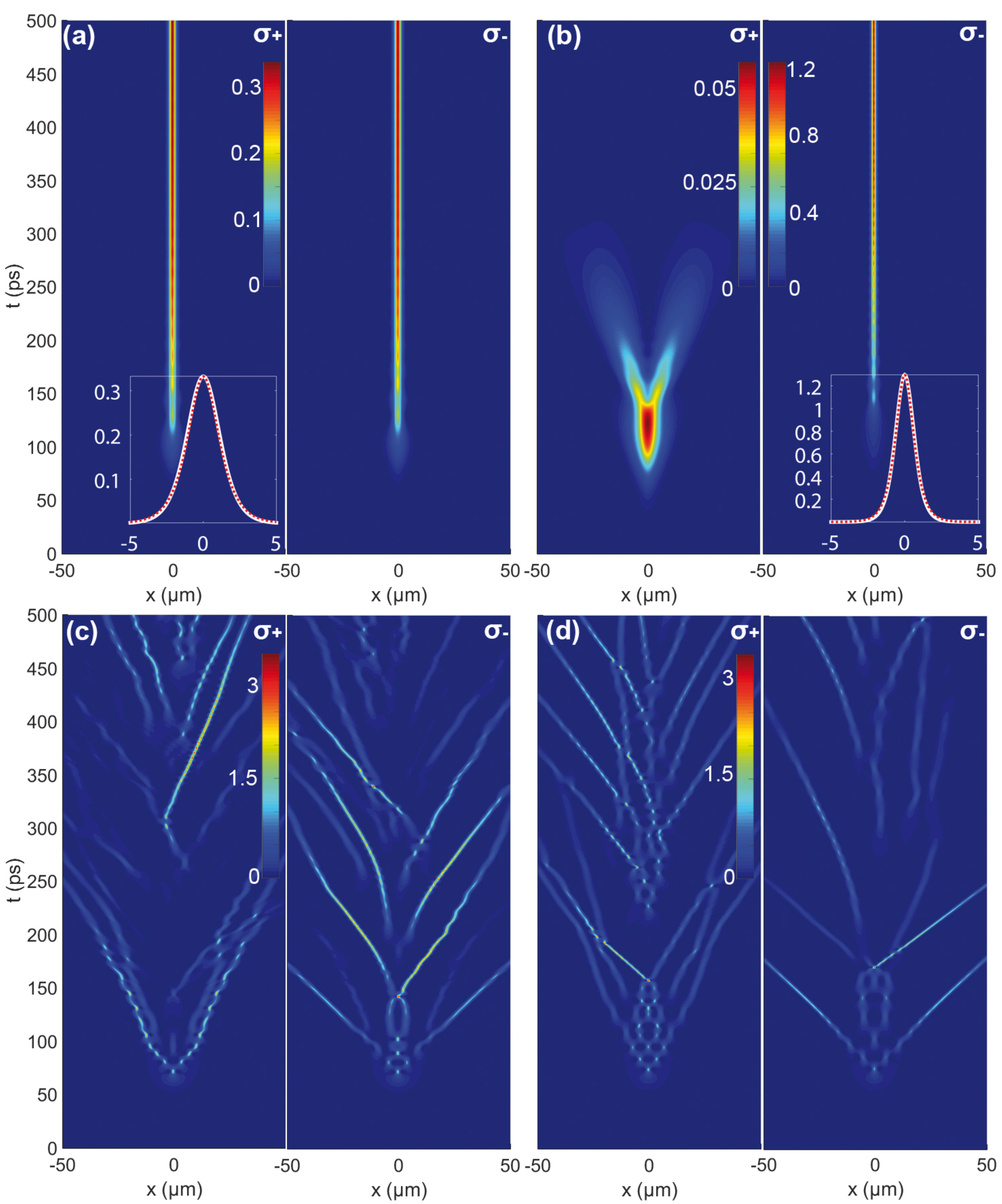}
\caption{Driven dissipative solitons, the colormap shows the chemical potential. The $\sigma_\pm$ label refers to the $\pm$ component. (a) Stable soliton solution obtained with $\sigma_P=2 \mu$m. (b) Stable half-soliton for $\alpha_2<0$. Unstable dynamics for $\sigma_P=15 \mu$m. (c) $\alpha_2<0$, (d) $\alpha_2>0$. The insets in panel (a) and (b) shows a slice at $t=500$ ps (white line) fitted by the sech analytical function (dashed-red line). In both cases the mean absolute error between the numerically and analytically generated curves is less than $10^{-5}$ meV.}
\label{Num3}
\end{figure}

Let us now treat the full problem as described by Eq. \ref{natural} including the effects of the particles being scattered into the condensate modes, i.e. formally $P\neq0$ and decay $\Gamma\neq0$. The initial population profile is set by the pump amplitude fixed to $A_P=250\Gamma_R$. A stable soliton solution can be obtained from a narrow pump spot where $\sigma_P=2 \mu m$ as shown in Fig. \ref{Num3} (a) where overlapping stationary bright solitons are formed after $100$ ps in both components. Interestingly the sustaining pump can wash out small perturbations in such a way that the soliton stability region is narrow but not singular. The inset shows a slice at $t=500 ps$ (white line) with the analytical sech (dashed-red line) from which we deduce a soliton width of $W=0.85$ $\mu m$.

Supposing the reservoir favors one component over the other, as it would be the case e.g. when pumping the condensate with a spin polarized pump laser or spontaneously in trapped condensates far away from the pumping spots \cite{Ohadi}, one can introduce a symmetry breaking between the two spin components. In Fig. \ref{Num3} (b) we have increased the transfer rates $\gamma$ to the $\sigma_+$ component by $1\%$ and considered $\alpha_2>0$. This scenario results in an imbalance between the $\sigma_\pm$ polariton/atoms populations and not only the $\sigma_+$ component does not match the stable bright soliton anymore leading to dominant dispersion but also it is reduced by the cross repulsion brought by $\alpha_2>0$. As a result the $\sigma_+$ population vanishes leaving a stable half-soliton in the $\sigma_-$ component of width $W=0.55$ $\mu$m (see inset), which is in accordance with the experimental findings in \cite{Ohadi}. However in \cite{Ohadi} spin selection happened to be spontaneous without any imbalance of the pump rates as discussed here. In the case $\alpha_2<0$ (not shown) the cross attraction prevents the spreading to occur leading to the overlapping stationary solitons.

Finally in the most common case where the pump spot size does not match the the optimal soliton condition as illustrated in Fig. \ref{Num3} (c) and (d), we obtain a very complex {\it fireworks dynamics} resulting from the multiple cross/self-interactions and competition and the breathing effects. Here we used the parameters $\sigma_P=15\mu m$ for the spot size in the case $\alpha_2<0$ in panel (c) and $\alpha_2 > 0$ in panel (d). We note that including a weak noise to the system dynamics, that is present in state-of-the-art experiments, leads to random patterns of the bright solitons, which can be utilized for the realization of a true random number generator.

\section{Conclusions} We have presented an approximate analytical treatment for the parameters of spin sensitive non-equilibrium bright solitons that is based on a non-conservative Lagrangian formalism. This Lagrangian formalism has been extended from a previous work to include the attractive interaction case of the non-equilibrium spinor condensate field and turned out to provide reliable results. Explicit ordinary parameter differential equations for the amplitudes, widths and phases were derived for the coherent and spin sensitive case and interpreted in terms of key system parameters. Particularly explicit expressions for a stationary solution of the bright soliton in a non-equilibrium environment were stated. In addition the scenario of non-equilibrium half-bright solitons was elucidated analytically and we found corresponding parameter equations and explicit static spinor expressions that include the coherent case as limit for vanishing cross-interactions and competition.  The explicit solutions allowed us to clarify the role of cross-interactions and interspecific competition between the two spin components of the condensate for the system parameters, the latter inducing a decrease of amplitudes. Furthermore we have demonstrated numerically the generation and high control over bright soliton emergence in non-equilibrium systems and highlighted several effects important for the fate of the bright solitons.  The existence of stationary equilibrium bright solitons in the spin coherent case as well as  breathing solutions with subsequent spontaneous symmetry breaking was shown, which is a simple manifestation of a true random number generator. In addition we considered the non-equilibrium dynamics of spinor condensates, where we have found for instance the novel phenomenon of bright soliton fireworks dynamics. Finally we have compared the analytical and numerical expressions and obtained a very good match of the presented results.


\[
\]

{\bf Ethics statement.} This work only discusses ethical computer simulations and mathematical analysis.

\[
\]

{\bf Data accessibility statement.} This work does not have any experimental data.

\[
\]

{\bf Competing interests statement.} We have no competing interests.

\[
\]

{\bf Authors contributions statement.} FP and HF conceived the mathematical models, interpreted the computational results, and wrote the paper. All authors gave final approval for publication.

\[
\]

{\bf Funding statement.} F.P. acknowledges financial support through his Schr\"odinger Fellowship at the University of Oxford and the NQIT project (EP/M013243/1).


 \end{document}